\documentclass{template/svproc}
\usepackage{float}
\usepackage{graphicx}
\usepackage{url}
\usepackage{hyperref}
\usepackage{xcolor}

\usepackage{fancyhdr}

\begin{document}
\mainmatter              
\title{Comprehensive Evaluation of CNN-Based Audio Tagging Models on Resource-Constrained Devices}
\titlerunning{Audio Tagging on Resource-Constrained Hardware}  
%
\author{Jordi Grau-Haro \and Ruben Ribes-Serrano \and Javier Narnajo-Alcazar \and
Marta Garcia-Ballesteros \and Pedro Zuccarello}
\authorrunning{Jordi Grau-Haro et al.} 
%
%
\institute{Instituto Tecnologico de Informatica (ITI), Valencia, Spain\\
\email{\{jgrau, rribes, jnaranjo, martagarcia, pzuccarello\}@iti.es}
}

\maketitle              

\begin{abstract}
Convolutional Neural Networks (CNNs) have demonstrated exceptional performance in audio tagging tasks. However, deploying these models on resource-constrained devices like the Raspberry Pi poses challenges related to computational efficiency and thermal management. In this paper, a comprehensive evaluation of multiple convolutional neural network (CNN) architectures for audio tagging on the Raspberry Pi is conducted, encompassing all 1D and 2D models from the Pretrained Audio Neural Networks (PANNs) framework, a ConvNeXt-based model adapted for audio classification, as well as MobileNetV3 architectures. In addition, two PANNs-derived networks, CNN9 and CNN13, recently proposed, are also evaluated. To enhance deployment efficiency and portability across diverse hardware platforms, all models are converted to the Open Neural Network Exchange (ONNX) format. Unlike previous works that focus on a single model, our analysis encompasses a broader range of architectures and involves continuous 24-hour inference sessions to assess performance stability. Our experiments reveal that, with appropriate model selection and optimization, it is possible to maintain consistent inference latency and manage thermal behavior effectively over extended periods. These findings provide valuable insights for deploying audio tagging models in real-world edge computing scenarios.
\end{abstract}
\section{Introduction}\label{sec:intro}

Edge computing for audio tagging has emerged as a critical enabler for a wide range of real-time, context-aware applications, especially where low latency, privacy preservation, and energy efficiency are essential. For instance, the work by Yang et al. \cite{yang2025real} presents a real-time acoustic scene recognition system deployed on edge devices to monitor daily routines of elderly individuals, aiming to improve assisted living environments through passive audio sensing. Additional research efforts in this area have also explored the monitoring of elderly people using ambient intelligence and privacy-preserving sensing techniques \cite{rashidi2012survey,wang2019privacy}. Lamrini et al. \cite{lamrini2023evaluating} evaluate pre-trained convolutional neural networks for anomaly detection in smart cities, leveraging embedded platforms to perform audio classification close to the source of the data. This approach minimizes data transmission needs while enabling timely detection of unusual events. Beyond these, edge-based audio tagging solutions are increasingly used in wildlife monitoring systems, where autonomous sensor nodes classify animal sounds to track species presence and behavior in remote areas \cite{stowell2019automatic}. Another prominent application is found in industrial sound monitoring, where local AI models are deployed on embedded hardware to detect mechanical failures or abnormal operating conditions in machinery, thus preventing downtime and optimizing maintenance schedules \cite{shaikh2023sound}. These examples underscore the increasing importance of deploying efficient and accurate audio classification models on edge devices, particularly in scenarios where cloud-based processing is impractical due to bandwidth, latency, or power constraints.

Previous studies, such as Bibbo et al. \cite{bibbo2023audio}, have explored the feasibility of implementing CNN-based audio tagging models on devices like the Raspberry Pi. Their work primarily focused on evaluating the performance of a single architecture, CNN9, highlighting challenges related to thermal management and inference latency over extended periods.

Building upon this foundation, our study presents a comprehensive evaluation of multiple CNN architectures for audio tagging on the Raspberry Pi. Specifically, we assess all models proposed in the Pretrained Audio Neural Networks (PANNs) framework \cite{kong2020panns}, encompassing both 1D and 2D CNN variants. PANNs have become a cornerstone in the field of computational audition, offering robust and transferable representations across a wide range of audio-related tasks. Trained on the extensive AudioSet dataset \cite{gemmeke2017audio}, which encompasses over 500 sound classes, PANNs have demonstrated state-of-the-art performance in audio tagging, surpassing previous benchmarks \cite{kong2020panns}. Beyond audio tagging, PANNs have been effectively applied to sound event detection (SED). For instance, PANNs served as a foundational component in a two-stage framework, significantly enhancing detection performance \cite{khandelwal2023leveraging}. Moreover, recent research has demonstrated the applicability of PANNs to other tasks beyond tagging and event detection, such as automated audio captioning. In this context, PANNs are typically employed as pretrained acoustic encoders, providing high-level semantic features that improve caption generation quality. Koh et al. \cite{koh2022automated} leveraged PANNs within a transfer learning framework and introduced a reconstruction latent space similarity regularization technique to enhance caption relevance. Similarly, Liu et al. \cite{liu2022leveraging} combined PANNs-derived embeddings with a BERT-based language model to generate more contextually accurate descriptions of audio content. Additionally, Xiao et al. \cite{xiao2023graph} incorporated PANNs features into a graph attention network to better capture temporal relationships in the acoustic signal. These studies underscore the versatility and robustness of PANNs across diverse audio understanding tasks, highlighting their relevance not only in classification but also in generative and sequence modeling domains. 

This wide range of possible applications highlights the importance of thoroughly understanding the performance of PANNs on resource-constrained devices. Given their versatility and strong generalization capabilities, evaluating how PANNs behave under the computational limitations of embedded platforms is crucial to enabling real-world deployment. Such analysis is especially relevant as edge computing becomes increasingly common in practical audio processing scenarios, where models must operate efficiently without sacrificing performance. Therefore, this work aims to provide valuable insights into the feasibility of deploying PANNs-based solutions in low-power environments, facilitating their integration into a broader spectrum of audio-related tasks.

Additionally, we incorporate the ConvNeXt-tiny model, adapted for audio tagging tasks as demonstrated by Pellegrini et al. \cite{pellegrini2023adapting}. Furthermore, we evaluate MobileNetV3 architectures proposed in \cite{schmid2023efficient,schmid2024dynamic}, as well as two convolutional networks following the PANNs design philosophy, namely CNN9 and CNN13, introduced in \cite{bibbo2023audio}. All these models have also been trained with Audioset.

By converting these models to the Open Neural Network Exchange (ONNX)\footnote{\url{https://github.com/onnx/onnx}} format, we ensure efficient deployment and portability across diverse hardware architectures.

A key differentiator of our work lies in the extensive evaluation period; each model is subjected to continuous inference tasks over a 24-hour duration. This approach allows us to monitor and analyze critical performance metrics, including inference time stability, CPU temperature fluctuations, and overall system reliability. Contrary to the observations reported in \cite{bibbo2023audio}, our findings indicate that, with appropriate model selection and optimization, it is possible to maintain consistent inference performance without significant thermal degradation over prolonged operation. In addition, inference was conducted under two different scenarios, depending on the presence of a graphical user interface. In the first scenario, the application performs inference and displays the results via the terminal (headless). In the second, the graphical interface is updated each time an inference is completed. This comparison enabled us to analyze the performance impact of using a user interface on the Raspberry Pi in real-time audio classification tasks.

Through this study, we aim to provide valuable insights into the practical considerations of deploying various CNN-based audio tagging models on edge devices, facilitating informed decisions for real-world applications in domains such as environmental monitoring, smart home systems, and assistive technologies. To the best of our knowledge, this is the first work to conduct such a comprehensive analysis of the performance of audio classification models on IoT or resource-constrained devices implementing models from the PANNs family and additional architectures explored in this study.

\section{Method}\label{sc:method}

\begin{figure}
    \centering
    \includegraphics[scale=0.35]{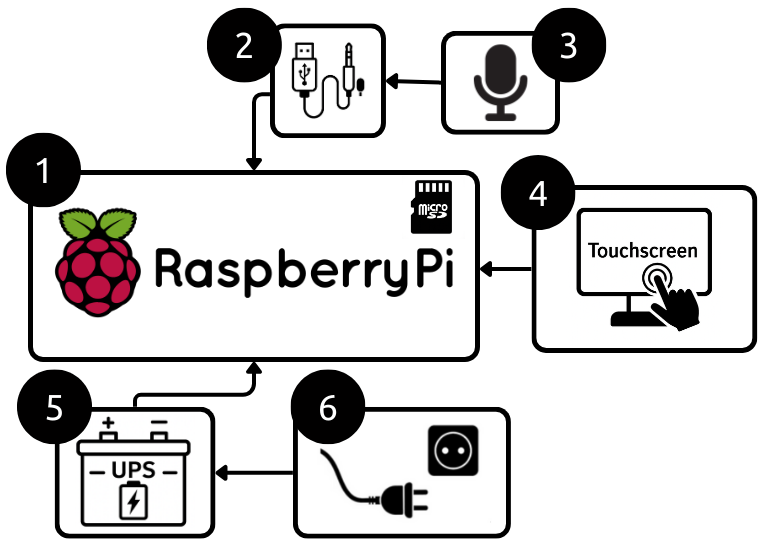}
    \caption{Block diagram of the experimental setup of the inference node}
    \label{fig:setup}
\end{figure}

This section outlines the methodology followed to conduct a comprehensive comparative analysis of various audio tagging models on an edge computing platform. First, we describe the hardware setup, detailing the Raspberry Pi device and its connected peripherals. Next, we explain the model conversion process, in which pre-trained neural networks are optimized for deployment on resource-constrained environments. Subsequently, we present the real-time inference application developed for this study, highlighting its design and integration with the audio pipeline. Finally, we provide experimental details, including the evaluation protocol, performance metrics, and testing conditions used throughout the analysis.

The inference device is composed of the following components:

\begin{enumerate}
    \item Raspberry Pi: the model used for the experiments corresponds to the Raspberry Pi 4B with 4 GB of RAM.
    \item External sound card: to enable audio input on the Raspberry Pi, an external USB sound card is employed. Specifically, the UGREEN USB 2.0 external sound card is used, which connects directly to the Raspberry Pi via a USB port. This setup ensures reliable audio capture and compatibility with a wide range of microphones, overcoming the limitations of the Raspberry Pi’s native audio interface.
    \item Microphone: the microphone used corresponds to the RODE Lavalier II model.
    \item Touchscreen: the inference GUI can be visualized by means of a 7'' LCD touch screen
    \item Unit Power Supply (UPS): to protect the Raspberry Pi from possible micro outages or complete shutdowns, it is connected to a UPS unit, specifically the PiJuice HAT model.
    \item Charger: the light charger is connected to the UPS.
\end{enumerate}


\subsection{Model Conversion}\label{subsec:model}

All models used in this study were converted from the original checkpoint files in Torch format as released by the authors. The design of the inference pipeline is based on the example implementations provided by the original authors of each model. Fig.~\ref{fig:model_mods} offers the reader a visual illustration that complements the upcoming explanations. All models were trained using audio sampled at 32 kHz. Consequently, when analyzing 10-second audio segments during each iteration, the input to the inference pipelines consistently consists of 320,000 samples. This choice aligns with the structure of the Audioset dataset, which is composed of 10-second clips, providing the models with a sufficient temporal context to capture relevant acoustic events. Further details on the real-time implementation can be found in Section~\ref{subsec:experiment}.

\subsubsection{PANNs and ConvNeXt-tiny}\label{subsubsec:panns_onnx}

: in contrast to the approach presented in \cite{bibbo2023audio}, which utilizes the native Torch model for inference, the present study adopts a different strategy by converting the models to the ONNX format. ONNX is widely recognized for its cross-platform compatibility and efficiency, particularly when deployed on resource-constrained CPUs. Consequently, all PANNs models\footnote{\url{https://github.com/qiuqiangkong/audioset_tagging_CNN/}} in this work were exported to the ONNX format with a fixed input size corresponding to 10 seconds of audio sampled at 32 kHz, resulting in an input tensor of 320,000 audio samples. ConvNeXt-tiny\footnote{\url{https://github.com/topel/audioset-convnext-inf/}} is converted following the same procedure.

\subsubsection{MobileNetV3}\label{subsubsec:mobilenet_onnx}

: following the official implementation provided by the authors\footnote{\url{https://github.com/fschmid56/EfficientAT}} as described in \cite{schmid2023efficient,schmid2024dynamic}, the inference process was divided into two separate models. The first model is responsible for spectrogram extraction, as the official implementation relies on a dedicated Torch module for this purpose. The resulting spectrograms are then fed into the second model, which consists of the MobileNetV3 architecture for audio tagging. Both models are converted to ONNX format. Consequently, the inference process is carried out in two stages: first, the audio input is processed by an initial model that converts it into a two-dimensional representation; then, this output is passed to a second model—based on MobileNetV3—which generates the final prediction.

\subsubsection{CNN9 and CNN13}\label{subsubsec:CNN9_onnx}

: in line with the work presented at \cite{bibbo2023audio}, it was also decided to include these two networks in the study. According to the official implementation\footnote{\url{https://github.com/gbibbo/ai4s-embedded}}, the spectrograms are computed using the Librosa\footnote{\url{https://librosa.org/doc/latest/index.html}} library and subsequently used as input for the model converted to the ONNX format. 

\begin{figure}[]
    \centering
    \includegraphics[width=\linewidth, ]{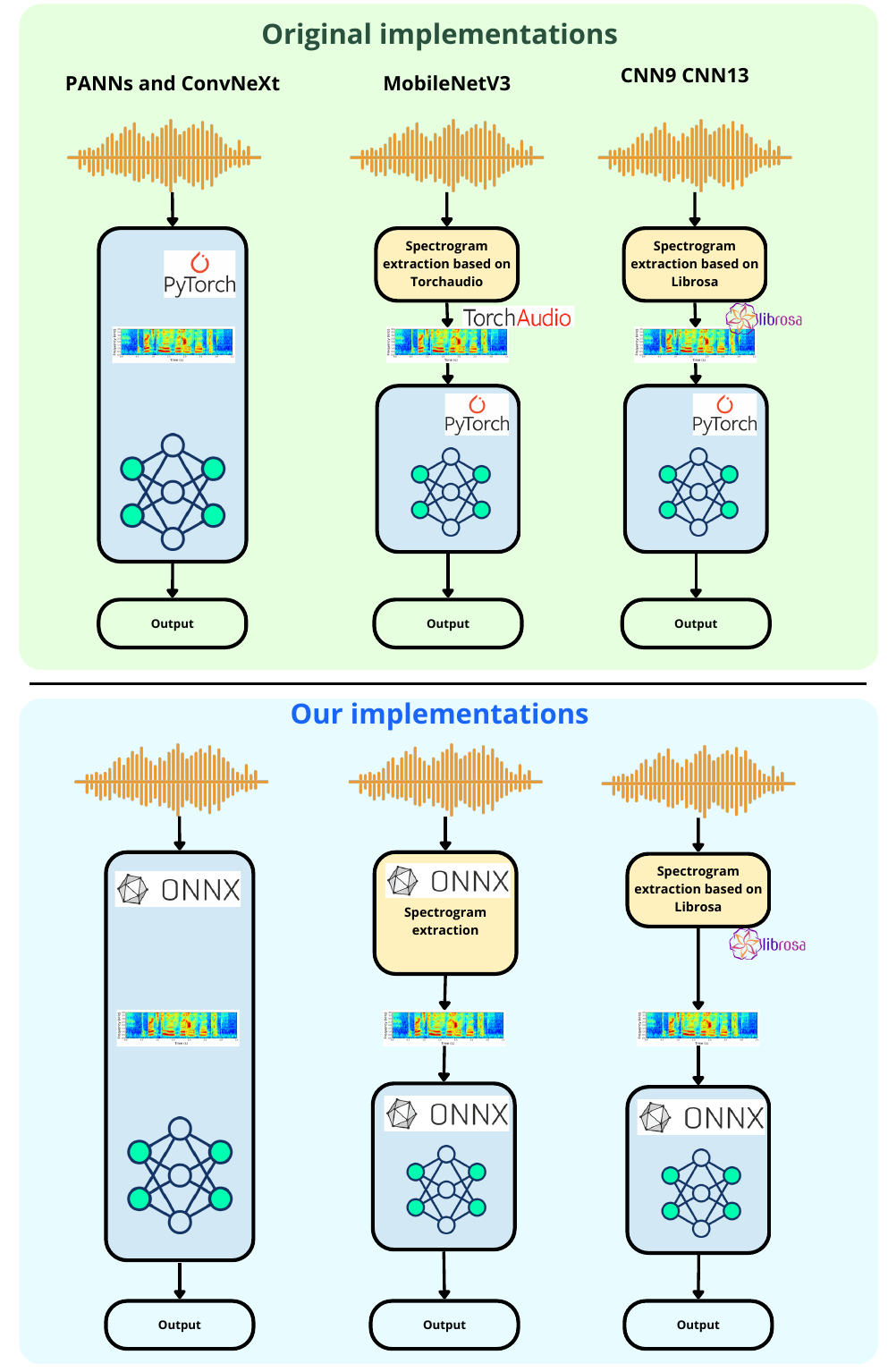}
    \caption{Comparison between the original and adapted implementations of audio classification models for deployment on Raspberry Pi. The top section shows the original PyTorch-based implementations for the PANNs and ConvNeXt, MobileNetV3, and CNN9/CNN13 models, each using different spectrogram extraction backends (internal, TorchAudio, or Librosa). The bottom section illustrates the modifications introduced in our work: all models are converted to ONNX format to enable efficient inference, with spectrogram extraction either embedded into the ONNX model or performed externally using Librosa prior to inference.}
    \label{fig:model_mods}
\end{figure}

\begin{figure}[t]
    \centering
    \includegraphics[width=1\linewidth]{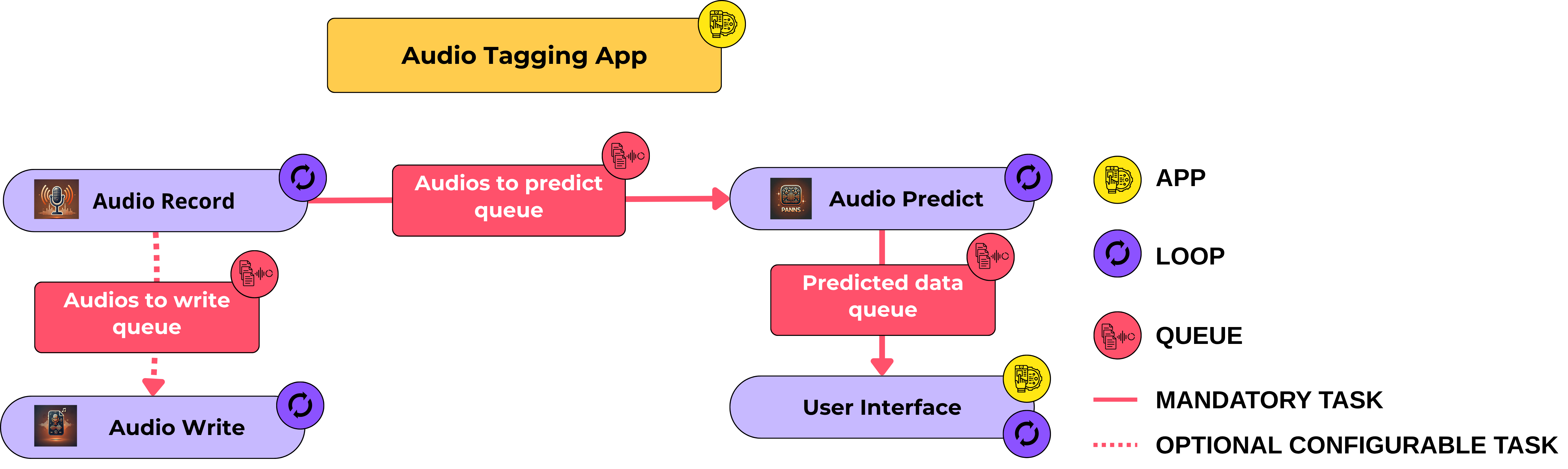}
    \caption{Diagram of the proposed edge audio tagging application}
    \label{fig:app}
\end{figure}

\subsection{Real Time Application}\label{subsec:app}

The system operates in a loop composed of four main modules: \textit{Audio Record}, \textit{Audio Write}, \textit{Audio Predict}, and \textit{User Interface}. For clarity, we describe the workflow corresponding to the scenario where the graphical user interface is active (see Fig.~\ref{fig:app}). It is important to note that, in the case where the experiment is conducted without the user interface, there is no communication between the \textit{Audio Predict} module and the \textit{User Interface}. Modules communicate via asynchronous queues (one queue per communication) to ensure real-time performance and modularity. Mandatory tasks (solid lines in Fig.~\ref{fig:app}) include recording the audio, predicting the sound class using a selected model (e.g., PANNs), and updating the user interface (when interface setup). Optional configurable tasks (dashed lines) such as saving the recorded audio to disk are also supported. This design enables flexible deployment on embedded systems by decoupling data acquisition, processing, and visualization, allowing the system to maintain responsiveness under constrained hardware resources. For the reader's convenience, Fig.~\ref{fig:app} is shown as a representation of the previous explanation. The purpose of each of the modules is explained below.

\begin{figure}[t]
    \centering
    \includegraphics[width=1\linewidth]{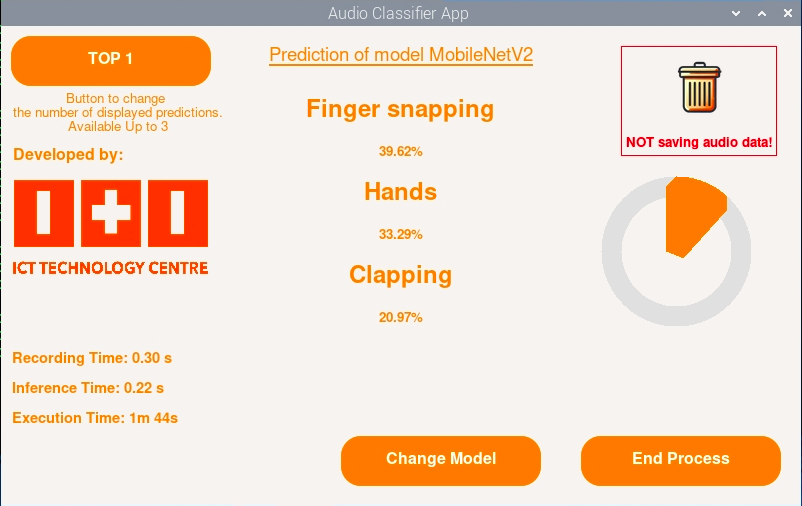}
    \caption{Graphical user interface (GUI) of the audio classification application running on the Raspberry Pi. The interface displays the top-3 predictions of the MobileNetV2 model for the captured audio segment, along with their associated confidence scores. Additional information, such as recording time (time since last prediction), inference time (time required to perform the last inference), and total execution time (time required to perform the last inference), is provided. The GUI allows the user to change the model or terminate the process, and explicitly informs that no audio data is being stored during inference.}
    \label{fig:gui}
\end{figure}

\begin{itemize}
    \item \textit{Audio Record}: this module performs audio recording. The audio is recorded with a sample rate of 41 kHz. Each time a 10-second audio is obtained, it is added to the prediction queue (see Fig.~\ref{fig:app} for the link between \textit{Audio Record} and \textit{Audio Predict}).
    \item \textit{Audio Write}: This optional module enables saving the audio previously captured by the \textit{Audio Record} module to disk. This functionality is particularly useful when there is an interest in storing audio samples for potential model refinement. In this setup, the \textit{Audio Record} module also adds each audio sample to a queue that is monitored by the \textit{Audio Write} module.
    \item \textit{Audio Predict}: This module is responsible for running inference on the audio data captured by the Audio Record module. It operates in a separate thread and continuously monitors a queue populated with audio segments. Each audio chunk is resampled to match the input shape expected by the ONNX model in order to obtain the prediction scores, which are sorted to identify the top classes along with their associated probabilities and inference time. The prediction results are placed into a dedicated queue for further processing or display. 
    \item \textit{User Interface}: The graphical interface provides a clear and intuitive visualization of the classification process. It displays the top audio class predictions returned by the model, along with their respective confidence percentages. The current model in use in Fig.~\ref{fig:gui} (e.g., MobileNetV2) is shown, and users can choose to display up to three prediction results by toggling the ``TOP 1" button. Key system metrics such as recording time, inference time, and total execution time are updated in real time. The interface also offers two main control buttons: ``Change Model” to switch the classification model, and ``End Process” to stop the application. A visual cue in the top-right corner indicates whether audio data is being saved, reinforcing transparency. The interface is designed for accessibility, with large fonts and a clear layout to support real-time monitoring and decision-making.
\end{itemize}


\subsection{Experimental Details}\label{subsec:experiment}

Each model runs continuously for 24 hours to perform inferences. Before launching the next model, the Raspberry Pi remains powered on but idle for 1 hour, performing no processing tasks. All experiments are conducted in a room with a stable ambient temperature of 25°C. The main metrics analyzed are the Raspberry Pi’s temperature and the inference time. Each audio segment analyzed has a duration of 10 seconds. To ensure temporal continuity and improve robustness, the system performs a prediction every 5 seconds using a sliding window approach. This means that each new prediction is based on the last 5 seconds of the previous segment and the next 5 seconds, resulting in a 50\% overlap between consecutive segments. This overlapping strategy creates a continuous loop of analysis, allowing the system to detect audio events more reliably over time.

\section{Results}\label{sc:results}

This section presents the results obtained from evaluating the performance of the models executed on the Raspberry Pi. The results correspond to 24-hour experiments, with the aim of analyzing the temporal evolution of CPU temperature (see Section~\ref{subsec:temperature}) and inference time (see Section~\ref{subsec:time}). The inference time refers to the duration required to generate a prediction following the processing pipeline of each model, as illustrated in Fig.~\ref{fig:model_mods}. Measurements, from each prediction every 5 seconds, were aggregated in 10-minute intervals to capture short-term variations and trends throughout the entire deployment. This approach allows for the identification of stability patterns, potential performance degradation over time, and the thermal impact of different model architectures and operating conditions. The results are presented comparatively using a figure composed of two subplots for the analyzed metric. The upper subplot (a) shows the system behavior in headless mode, while the lower subplot (b) illustrates the performance when the graphical user interface is enabled.


\subsection{Inference Time Analysis}\label{subsec:time}

\begin{figure}[t]
    \centering

    \begin{minipage}[b]{\linewidth}
        \centering
        \includegraphics[width=\linewidth]{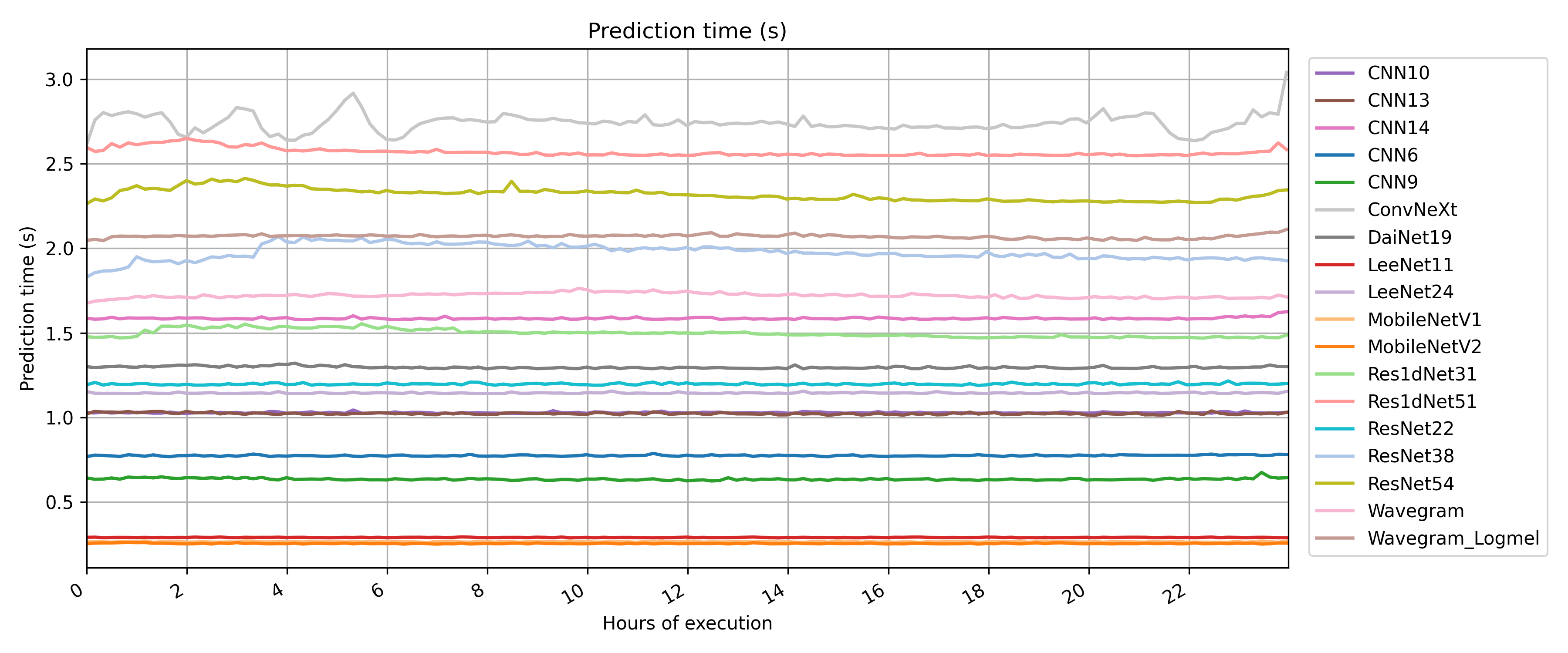}
        
        (a)
    \end{minipage}


    \begin{minipage}[b]{\linewidth}
        \centering
        \includegraphics[width=\linewidth]{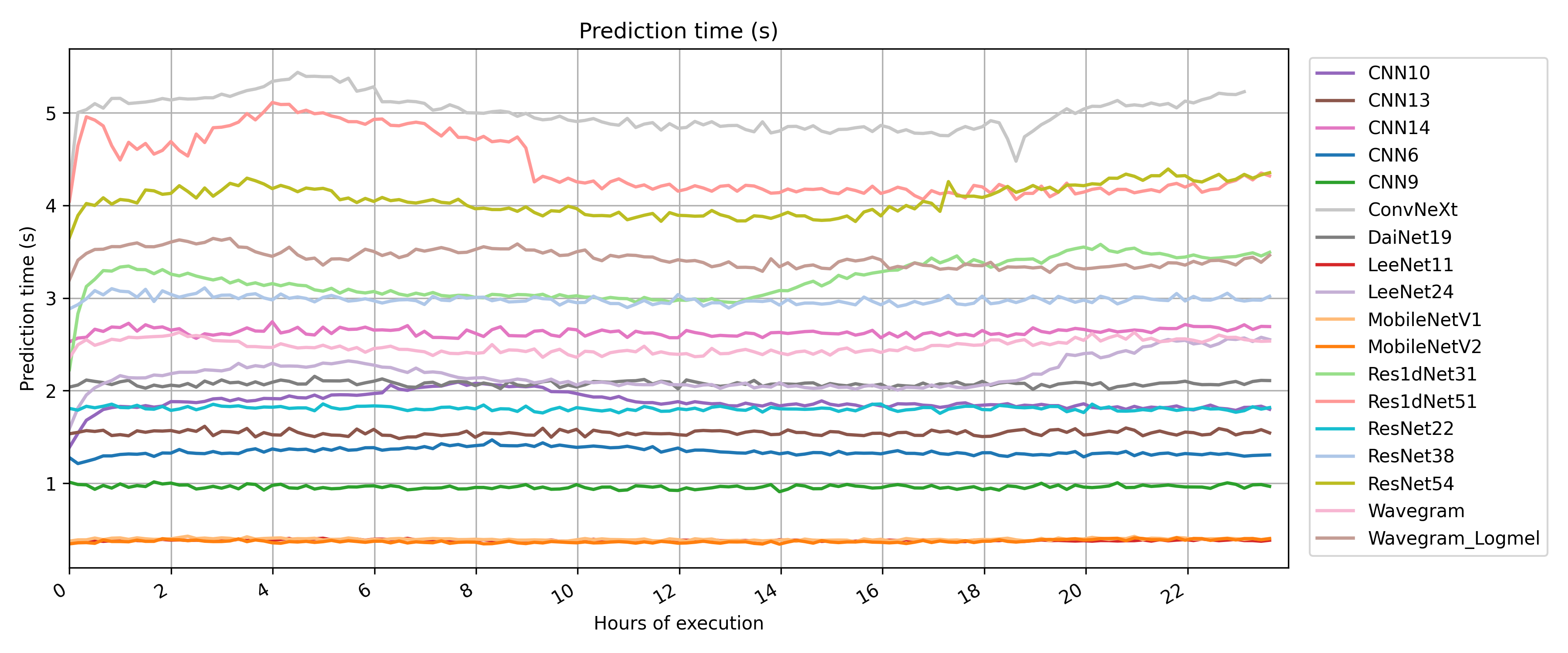}
        
        (b)
    \end{minipage}
    
    \caption{Inference time over 24 hours for different PANNs configurations. (a) System running headless (no user interface). (b) System with a real-time user interface.}
    \label{fig:pred_time_panns}
\end{figure}

\begin{figure}[t]
    \centering
    \begin{minipage}[b]{\linewidth}
        \centering
        \includegraphics[width=\linewidth]{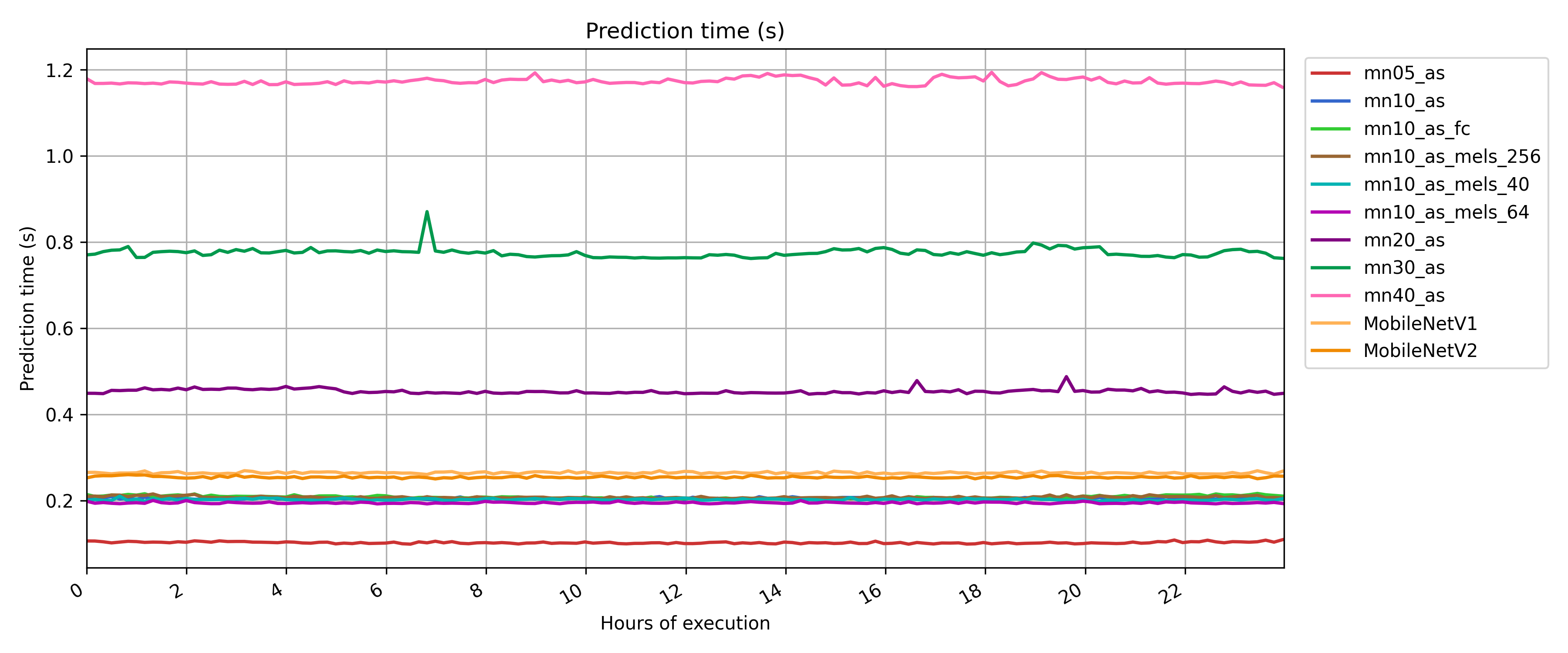}
        
        (a)
    \end{minipage}
    
    
    \begin{minipage}[b]{\linewidth}
        \centering
        \includegraphics[width=\linewidth]{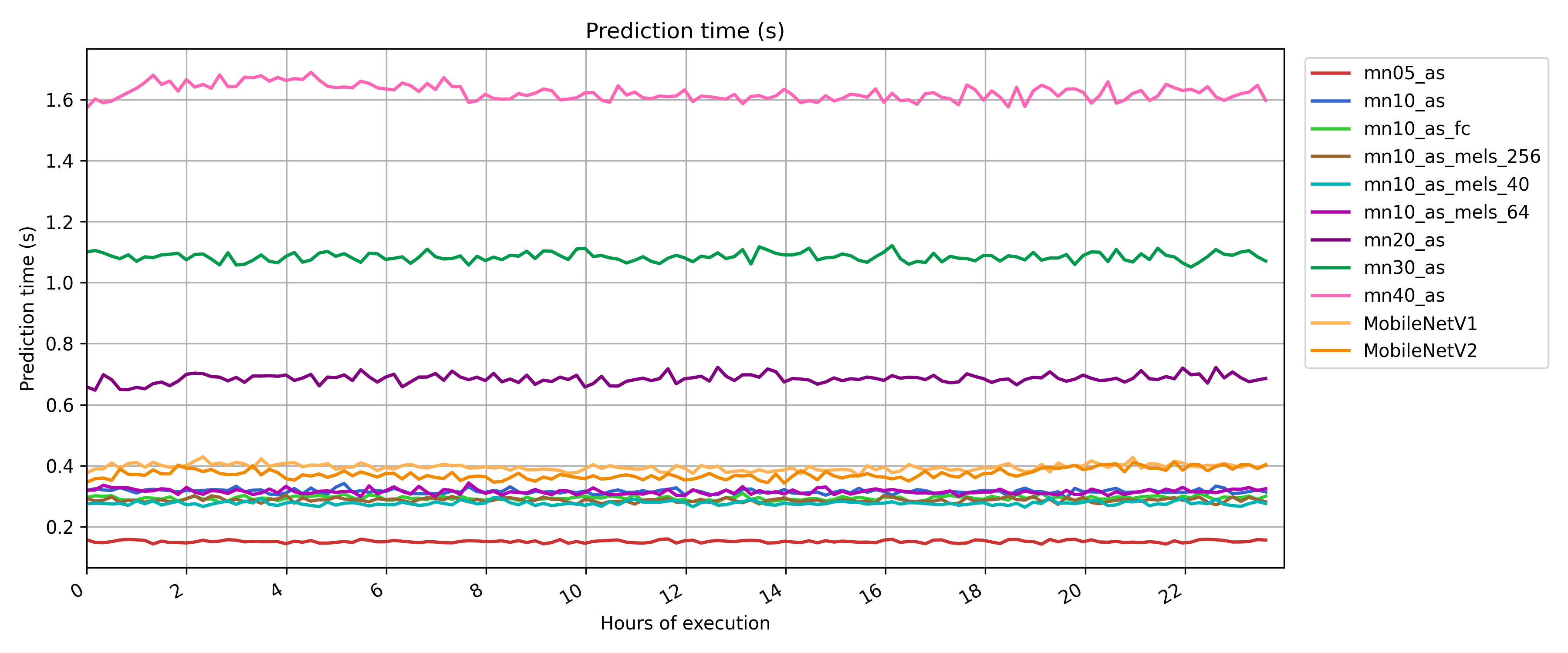}
        
        (b)
    \end{minipage}
    
    \caption{Inference time over 24 hours for different MobileNet configurations. (a) System running headless (no user interface). (b) System with a real-time user interface.}
    \label{fig:pred_time_mobilenets}
\end{figure}

\subsubsection{PANNs, ConvNeXt-tiny, CNN9 and CNN13}\label{subsubsec:time_panns}


Fig.~\ref{fig:pred_time_panns} presents the inference times of various convolutional neural network architectures over 24 hours of continuous execution. This analysis also includes the MobileNetV1 and MobileNetV2 models, as they are part of the PANNs family. However, a more detailed evaluation of these architectures is provided later, alongside the rest of the MobileNet variants. In the headless scenario, most models maintain stable and relatively low inference times. The fastest models include CNN9, CNN13, and CNN6, all showing consistent inference times under or around 1 second. In contrast, more complex architectures such as ConvNeXt, Res1dNet51, Wavegram\_Logmel, and ResNet54 exhibit higher inference times, generally between 2 and 3 seconds, though still stable over time.

When the GUI is active Fig.~\ref{fig:pred_time_panns}-(b), a general increase in inference time is observed across all models, along with higher temporal variability. Notably, ConvNeXt, ResNet54, and Res1dNet51 exhibit peaks exceeding 4 and 5 seconds, respectively, which could compromise their usability in real-time inference scenarios. The CNN14 network (widely used in state-of-the-art applications as it shows the best performance on Audioset) remains stable at around 3.5 seconds (2 seconds slower when compared to the headless scenario). The performance degradation appears more pronounced in deeper or more computationally intensive architectures, suggesting that the GUI introduces a system load that affects models unevenly.

These findings highlight the importance of considering the deployment environment when selecting a model for real-time applications. While some models perform well under controlled, headless conditions, their inference efficiency can deteriorate significantly with additional system overhead introduced by graphical interfaces. Therefore, the final choice of architecture should not only be based on baseline inference speed but also on its robustness to system load and operational conditions.



\subsubsection{MobileNets}\label{subsubsec:time_mobilenets}




Figure~\ref{fig:pred_time_mobilenets} presents the inference time over a 24-hour period for various MobileNet configurations.

A clear difference in performance is observed between the two scenarios. In the headless setup Figure~\ref{fig:pred_time_mobilenets}-(a), inference times are consistently lower and more stable across all configurations. Models such as mn05\_as, mn10\_as, and mn10\_as\_mels\_256 maintain inference times below 0.25 seconds with minimal variability. Even the heavier configurations, like mn30\_as and mn40\_as, show relatively stable behavior.

When the real-time interface is enabled, Figure~\ref{fig:pred_time_mobilenets}-(b), overall inference times increase, especially in the more demanding configurations. For instance, mn40\_as and mn30\_as show more pronounced fluctuations and reach higher peaks (up to 1.6 seconds for mn40\_as). Even the lighter models experience a slight upward shift in latency, though they remain relatively stable.

This comparison highlights the computational overhead introduced by real-time visualization tasks. While the system remains functional under both conditions, the headless mode clearly offers superior timing consistency—particularly important in low-latency or resource-constrained deployments.

\begin{figure}[t]
    \centering
    \begin{minipage}[b]{\linewidth}
        \centering
        \includegraphics[width=\linewidth]{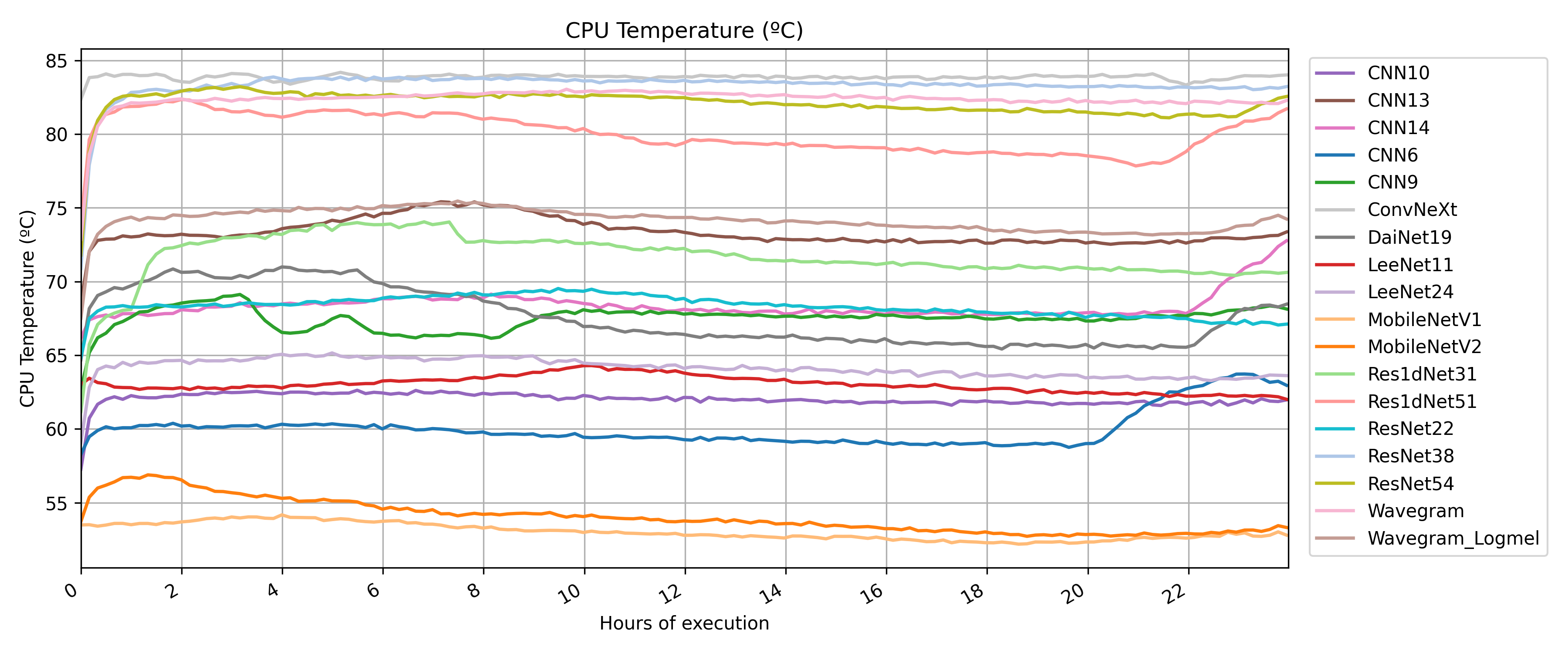}
        
        (a)
    \end{minipage}
    
    
    \begin{minipage}[b]{\linewidth}
        \centering
        \includegraphics[width=\linewidth]{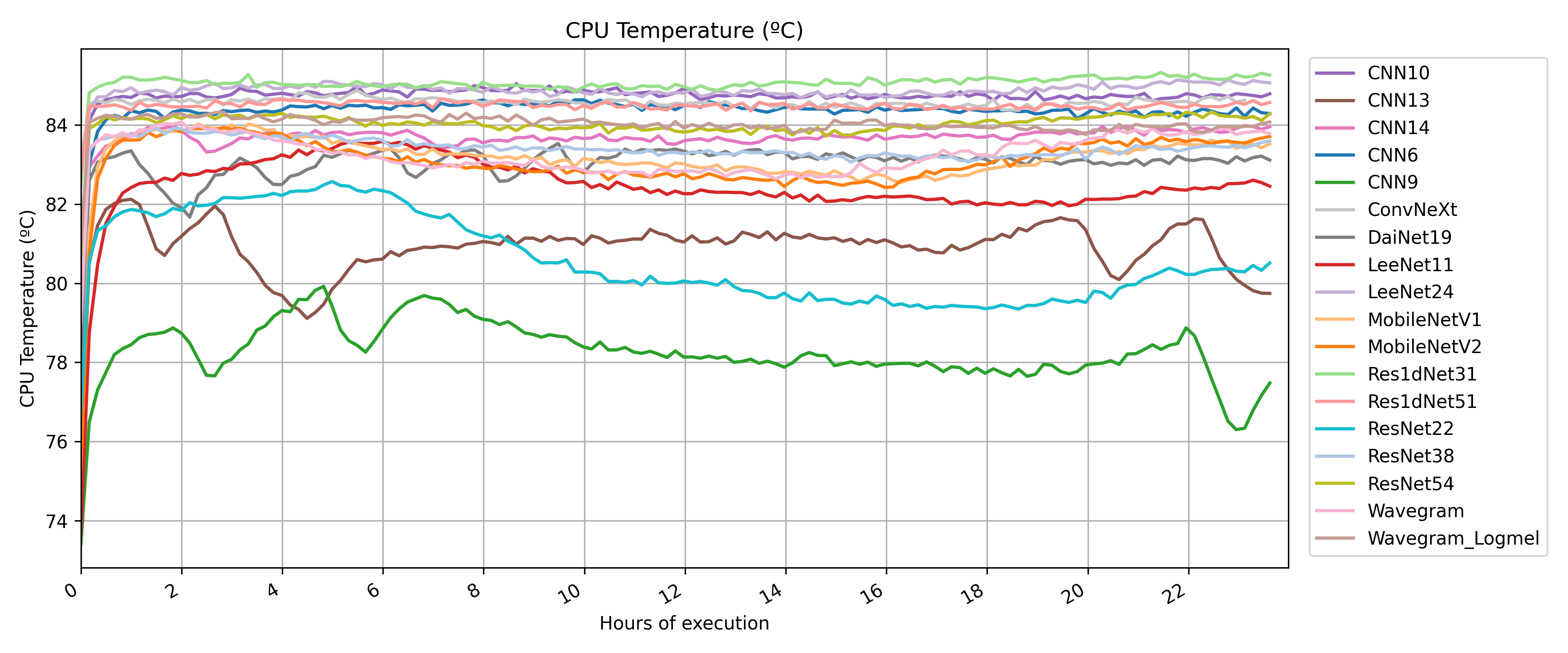}
        
        (b)
    \end{minipage}
    
    \caption{CPU temperature during 24 hours of continuous inference using different PANNs configurations. (a) Inference-only execution. (b) Inference with a real-time graphical user interface.}
    \label{fig:temp_panns}
\end{figure}

\subsection{Temperature Analysis}\label{subsec:temperature}

\subsubsection{PANNs, ConvNeXt-tiny, CNN9 and CNN13}\label{subsubsec:temp_panns}


Fig.~\ref{fig:temp_panns} shows the CPU temperature evolution during 24 hours of continuous inference for various convolutional neural network models. In the headless setup, there is a clear differentiation in thermal behavior between models. Lightweight architectures such as MobileNetV2, CNN6, and MobileNetV1 exhibit the lowest and most stable temperatures, maintaining values well below 65~°C throughout execution. Conversely, heavier models like ResNet54, ConvNeXt, and Wavegram reach temperatures above 80~°C, indicating higher computational demands.

In contrast, when the GUI is active Fig.~\ref{fig:temp_panns}-(b), nearly all models converge to higher and more uniform temperature values, clustering around 83–85~°C. This effect is particularly evident for models that previously had moderate temperatures in the headless setup, such as CNN6, CNN10, and CNN14, which show a significant increase. Moreover, temperature stability decreases under GUI conditions, with some models—like ResNet22 and CNN9—exhibiting fluctuations and transient drops that are not present in the headless case.

Although CNN9 and CNN13 exhibit fluctuating and unstable inference times over the 24-hour period, they do not cause a significant increase in system temperature when the GUI is active. As shown in Fig.~\ref{fig:temp_panns}-(b), all PANNs models—except for ResNet22—reach critically high temperatures, which suggests that their deployment may not be optimal for this scenario (see Fig.~\ref{fig:model_mods} for implementation explanation).

These results emphasize the influence of system load and graphical overhead on thermal behavior. While certain models may appear thermally efficient in headless operation, their performance can degrade under realistic deployment scenarios involving user interfaces. Importantly, some models (e.g., CNN14) consistently reach critical temperatures close to or above 85 °C, which may lead to thermal throttling or hardware degradation if used long-term without proper cooling. Therefore, temperature profiling under deployment-specific conditions is essential when selecting models for real-time embedded or resource-constrained applications.

\begin{figure}[t]
    \centering
    \begin{minipage}[b]{\linewidth}
        \centering
        \includegraphics[width=\linewidth]{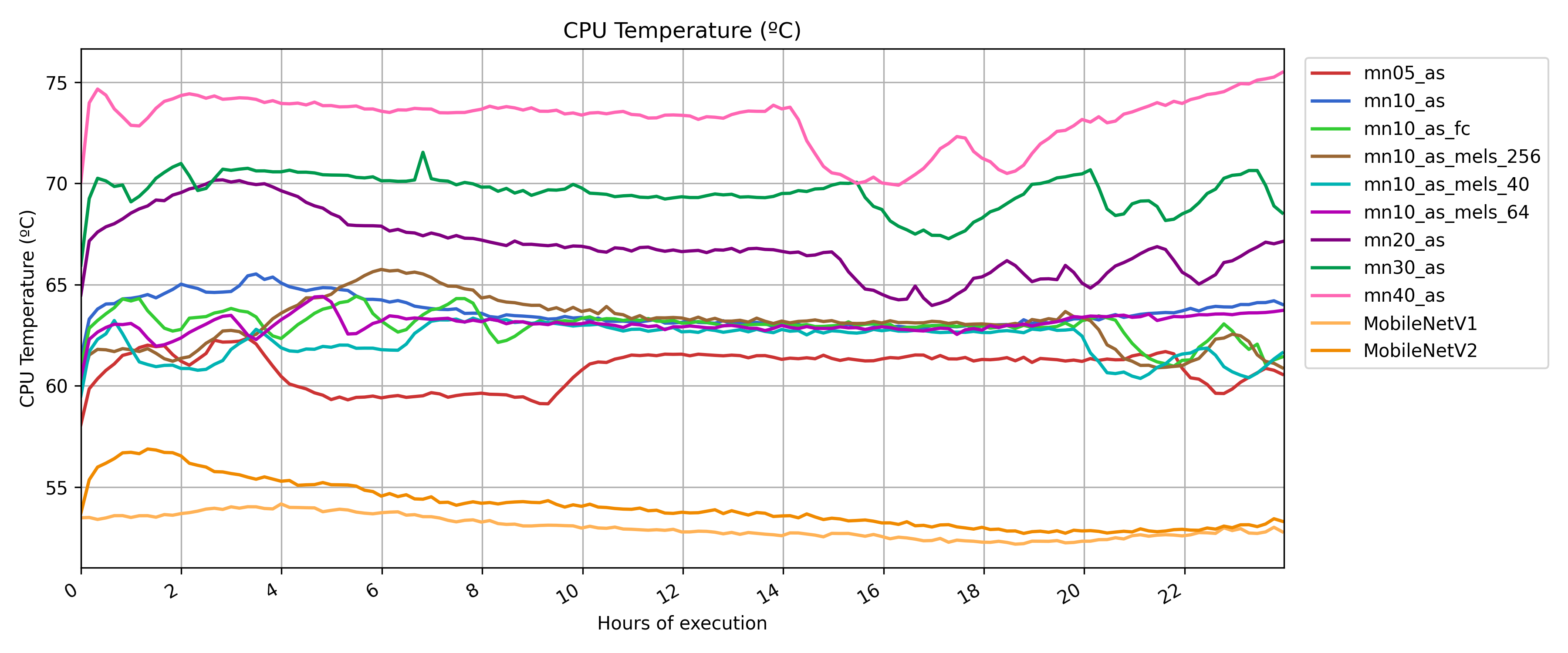}
        
        (a)
    \end{minipage}
    
    
    \begin{minipage}[b]{\linewidth}
        \centering
        \includegraphics[width=\linewidth]{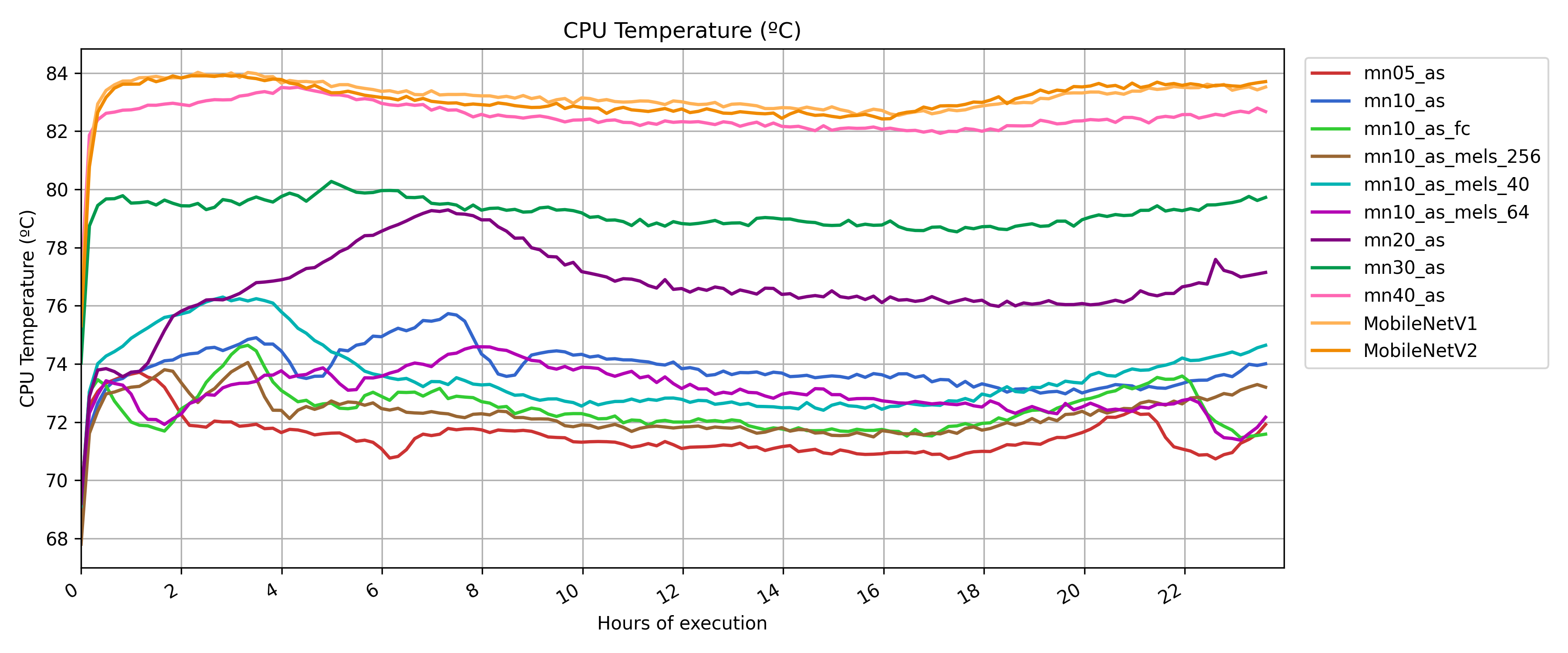}
        
        (b)
    \end{minipage}
    
    \caption{CPU temperature during 24 hours of continuous inference using different MobileNet configurations. (a) Inference-only execution. (b) Inference with a real-time graphical user interface.}
    \label{fig:temp_mobilenets}
\end{figure}

\subsubsection{MobileNets}\label{subsubsec:temp_mobilenets}

Figure~\ref{fig:temp_mobilenets} presents the evolution of CPU temperature during 24 hours of continuous execution for different MobileNet configurations. Subplot Figure~\ref{fig:temp_mobilenets}-(a) corresponds to the headless mode. In this condition, CPU temperatures remain moderate, generally ranging between 55 ºC and 75 ºC depending on model complexity. Lightweight variants such as mn05\_as, mn10\_as, and MobileNetV1/V2 maintain particularly low and stable temperatures, indicating efficient thermal behavior under minimal system load.

In subplot Figure~\ref{fig:temp_mobilenets}-(b), the additional processing overhead introduced by the GUI results in a noticeable temperature increase across all configurations. Models such as mn40\_as, MobileNetV1 and MobileNetV2 reach values above 82 ºC, while others show a consistent rise of approximately 5–10 °C compared to the headless setting. This highlights the significant impact of interface-related computational load on thermal behavior in embedded systems. The MobileNet models implemented within the PANNs framework (MobileNetV1 and MobileNetV2) transition from being the most thermally efficient in headless mode to the ones that generate the highest system temperatures when the GUI is enabled. This further reinforces the notion that the original PANNs implementation is not well-suited for deployment in scenarios involving a real-time graphical interface.

Furthermore, the data reveals that configurations such as mn20\_as, mn30\_as, and mn40\_as not only exhibit higher temperatures but also more variability over time, suggesting potential thermal instability or frequent load fluctuations. These findings emphasize the importance of considering both model complexity and system-level interactions (e.g., UI processes) when designing edge AI solutions, particularly in long-term deployments. If left unmitigated, sustained high temperatures could lead to thermal throttling or hardware degradation, especially in sealed enclosures with limited airflow. For practical deployments, thermal profiling under realistic workload scenarios becomes essential to ensure system reliability and energy efficiency.

\section{Disccussion}\label{sec:discussion}

This study demonstrates the feasibility of deploying MobileNet-based models for continuous real-time audio inference on low-power embedded devices. Among the evaluated configurations, MobileNetV3 variants provide a favorable balance between performance and computational efficiency, with several lightweight models (e.g., mn05\_as, mn10\_as) maintaining low inference times and stable CPU temperatures over extended periods of operation.

However, a key finding of this work is the significant thermal and computational impact introduced by the presence of a real-time graphical user interface. When running the same models under GUI conditions, CPU temperatures increased by up to 10 ºC across all configurations, with some models (notably mn40\_as, MobileNetV1 and MobileNetV2) reaching critical levels above 82 ºC. Such conditions may pose long-term risks to system stability and hardware integrity, especially in sealed environments with limited cooling capacity.

These results highlight the importance of considering the full deployment context—including interface requirements—when selecting and validating neural network models for edge applications. Model selection, even in models designed for low resources such as MobileNet, should not only optimize for inference accuracy and latency, but also account for thermal constraints and the potential overhead introduced by concurrent system tasks. For robust and safe deployments, especially in unattended or industrial environments, thermal profiling and system monitoring are essential components of the design and evaluation process.

Regarding PANNs family, based on the analysis of thermal behavior and prediction times, several conclusions can be drawn regarding the PANNs (Pretrained Audio Neural Networks) models, particularly CNN6, CNN10, CNN14, Wavegram, and Wavegram\_Logmel (proposed models in \cite{kong2020panns}).

First, PANNs models exhibit a wide range of computational demands, both in terms of CPU temperature and inference time. Among them, CNN6 stands out for its low computational load, maintaining low and stable temperatures in both headless and GUI scenarios, and achieving relatively short prediction times. This makes it a good candidate for real-time applications on resource-constrained systems. On the other hand, CNN14 and Wavegram\_Logmel consistently show high CPU usage and long prediction times, reaching critical temperatures that could jeopardize hardware integrity in prolonged execution scenarios. 

Second, the presence of a graphical user interface significantly affects PANNs models, especially the heavier ones. While CNN10 and CNN6 scale reasonably with the added system load, CNN14 and Wavegram\_Logmel suffer notable thermal stress, reaching the upper limits of safe CPU temperatures (~85 °C). This highlights the importance of considering deployment context—models that are viable in a headless setup may become impractical when GUI components are introduced. Wavegram demonstrates promising inference times—approximately 1.5 seconds in headless mode and 2.5 seconds with the user interface enabled. However, this performance comes at the cost of elevated CPU temperatures, reaching over 83 °C even in headless conditions.

The implementation of PANNs models exhibits critical thermal behavior when the graphical user interface is enabled. Among all evaluated configurations, it is the one that experiences the most significant increase in system temperature when transitioning from a headless environment to a GUI-based setup. When comparing the performance of PANNs models (e.g., CNN6, CNN10, CNN14, Wavegram, and Wavegram\_Logmel) with the full MobileNet family, including MobileNetV1, MobileNetV2, and the MobileNetV3 variants, several key differences emerge in terms of efficiency and thermal behavior. The PANNs architectures, particularly Wavegram\_Logmel and CNN14, consistently exhibit higher inference times and CPU temperatures, which may become critical under prolonged execution or in systems with limited cooling capabilities. This makes them less suitable for real-time or resource-constrained environments, despite their robustness in audio feature representation.

In conclusion, while PANNs offer excellent audio representation capabilities, their deployment should be carefully planned based on system constraints. For applications requiring continuous inference or embedded systems without active cooling, lighter models like CNN6 or CNN10 are preferable. Heavier models such as CNN14 and Wavegram\_Logmel, despite their potential in performance, pose risks due to high thermal output and should be used only when system cooling is not a limiting factor.




\section{Conclusion and Future Work}\label{sec:conclusion}

In this work, we conducted a comprehensive evaluation of a wide range of state-of-the-art audio tagging models, including MobileNetV1, V2, V3 (with different configurations), and PANNs, deployed on a Raspberry Pi under two operating conditions: headless and with a graphical user interface. The experiments assessed inference time, CPU temperature, and system stability over extended periods. Results highlight that while some lightweight models like MobileNetV2 and certain configurations of MobileNetV3 are suitable for real-time inference, others—particularly larger PANNs or unstable MobileNetV3 variants—can exceed safe thermal thresholds or show latency unsuitable for edge applications. As future work, the inference pipeline of the PANNs models will be restructured with the aim of enhancing their performance and thermal efficiency in scenarios where a graphical user interface is active. Furthermore, future work will aim to expand this experiment by incorporating additional IoT devices, such as NVIDIA Jetson platforms, or by integrating specialized AI hardware to enhance performance and scalability.

To the best of our knowledge, this is the first study with such scope and depth evaluating the practical deployment of audio tagging models on low-power edge devices. Previous efforts, such as Pellegrini et al.~\cite{bibbo2023audio}, focused on a single model and did not achieve thermal stability. Our work not only benchmarks performance across architectures but also demonstrates the importance of evaluating models under realistic deployment conditions, including the presence of a GUI. These insights are essential for designing reliable and safe real-time acoustic sensing systems.

\section*{Acknowledgements}

The participation of all the researchers in this work was funded by the Valencian Institute for Business Competitiveness (IVACE). The research carried out for this publication has been partially funded by the project STARRING-NEURO (PID2022-137048OA-C44) funded by the Agencia Estatal de Investigación (Spanish State Research Agency) and the Europan Union \\ MCIN/AEI/10.13039/501100011033/ FEDER, UE.

%
%

\bibliographystyle{template/bibtex/splncs03}  
\bibliography{template/referencias}    








\end{document}